\documentclass[aps,reprint,showpacs,showkeys,floatfix,longbibliography,superscriptaddress,prstper]{revtex4-1}

\usepackage{graphicx}
\usepackage{dcolumn}
\usepackage{amsmath}
\usepackage{amssymb}
\usepackage{epstopdf}
\usepackage{rotating}
\usepackage{color}
\usepackage{natbib}
\usepackage{url}

\begin{document}

\title{Are inequities in self-efficacy a systemic feature of physics education?}


\author{Jayson M. Nissen}\affiliation{Department of Science Education, CSU Chico, Chico, CA 95929} 

\begin{abstract}
There is consistent and growing evidence that  physics instruction disproportionately harms female students' self-efficacy, their beliefs about their ability to learn and do physics. This harm is problematic because self-efficacy supports student learning and persistence. Nissen and Shemwell (PhysRevPER, \textbf{12}, 2016) investigated this harm using an in-the-moment measure of student's self-efficacy states, which are dynamic judgments of one's ability to succeed in the activity at hand. Their results indicated that female students experienced much lower self-efficacy states in physics than male students did, and that this gender difference did not occur in other STEM courses. A limitation of their study was that it only investigated  a single college physics course. In order to further inform the generalizability of this phenomenon I analyzed a large data set of 35,464 experiences from 4,816 students at 33 secondary schools collected between 1993 and 1997 that was designed to be representative of high school students' experiences. Results confirmed that there was a large gender difference in self-efficacy states in high school physics courses and not in any other courses. The identification of this phenomenon in two very different settings indicates that physics instruction systemically harms female students' belief in their ability to learn and do physics.
\end{abstract}

\maketitle
\section{Introduction}
Self-efficacy is the belief in one's ability to succeed in a given task or domain \citep{Bandura1997}. In university physics courses self-efficacy predicts students' achievement \citep{Kost-smith2011,Sawtelle2012b}. This relationship in physics aligns with the broader findings on self-efficacy being consistently related to important academic and career outcomes. Increased self-efficacy \textit{causes} improved cognitive performance \citep{Bouffard1990}, increased interest in pursuing science degrees \cite{Luzzo1999}, and improved academic outcomes \cite{Williams2010}. Self-efficacy also predicts students taking harder courses \citep{Betz1983}, academic success \citep{Zajacova2005, Pietsch2003,Pajares1996}, the college major students choose \citep{Betz1983,Marra2009} and students' choice of career \citep{Bandura2001,Lent1993,Brown1989}.  
\par 	Given these relationships, a reasonable goal for physics education is to support students in developing self-efficacy beliefs. Yet, students' self-efficacy tends to decrease \citep{Kost-smith2011,Sawtelle2010,Lindstrom2011,Dou2016,Nissen2016} or at best not change \citep{Sawtelle2010,Lindstrom2011,Cavallo2004} from pre- to postinstruction in introductory physics courses. And this decrease is much larger for female students than for their male peers \cite{Kost-smith2011,Sawtelle2010,Lindstrom2011}. In contrast, students' self-efficacy increases from pre- to postinstruction in introductory chemistry courses \citep{Dalgety2006,Villafane2014,Ferrell2015}, an introductory algebra course \citep{Brewer2009} and introductory biology courses \citep{Lawson2007,Roster2006,Ainscough2016}. These differences indicate that the negative shift in self-efficacy is specific to female students in physics.

\section{Investigating Students' Self-Efficacy State Experiences in Physics}
The negative shift in female students' physics self-efficacy compared to the neutral shifts in male students' physics self-efficacy and the positive shifts for all students in other STEM domains indicates that the physics-learning environment may harm female students' physics self-efficacy. However, because all of these studies on self-efficacy in college STEM courses have only focused on one domain, they have not been able establish the physics-learning environment as causing the harm to female students' physics self-efficacy. For example, these studies leave open the possibility that male and female students have similar experiences in the physics-learning environment but reflect differently on these experiences, which is known as a state-trait discrepancy. To investigate if the physics learning environment harms female students' physics self-efficacy, Nissen and Shemwell \cite{Nissen2016} differentiated between two types of self-efficacy: states, which are dynamic and momentary, and traits, which are more stable attitudes. They used an in-the-moment sampling technique, the Experience Sampling Method, to measure college students' self-efficacy states throughout their daily lives. This method allowed them to compare male and female student's self-efficacy states in physics within the context of the students' self-efficacy states in other STEM courses and non-school activities. They found that female students experienced much lower self-efficacy states in physics than male students did; this difference was very large and female students self-efficacy states in physics were amongst the worst self-efficacy states that they experienced. In contrast, male and female students had similar self-efficacy states in other STEM courses, and those states were similar to male students' self-efficacy state in physics. Given that female students in the course had larger decreases in their self-efficacy traits than their male peers, as measured by pre/post surveys, and the large and unique difference in self-efficacy states in physics, they concluded that the physics-learning environment harmed female students' self-efficacy.
\par 	A limitation of Nissen and Shemwell's investigation was that it focused on a single physics course. They found evidence that this course was representative of research-based physics courses. For example, the course they investigated had similar student outcomes to other research-based physics courses for student grades, conceptual knowledge, attitudes, and self-efficacy beliefs.  This evidence of the generalizability of their findings indicated that most college physics-learning environments harm female students' self-efficacy. Nonetheless, the limitation of the study to one course and the recruitment of the students from that course leaves open the possibility that the findings in that study are not representative of a trend common to physics instruction. In order to determine the extent to which the harm to female students' self-efficacy is a systemic feature of physics instruction it is necessary to investigate gender differences in self-efficacy states and traits in a large and representative sample across many different domains.
\section{Purpose}
My purpose in this study was to inform the extent to which the physics-learning environment harms female students' physics self-efficacy is a systemic feature of physics instruction. This purpose is motivated by the weight of the evidence indicating that this harm is a common phenomenon in physics courses, and only physics courses, and the lack of evidence coming from a broad and representative sample. To meet this purpose I investigated the extent to which the gender difference in self-efficacy states in physics occurs in a large and representative sample of high school students. This data represents a very different reference point than the data collected by Nissen and Shemwell \citep{Nissen2016} in the research-based university physics course because it is drawn from a national scale investigation of secondary students that was not focused on physics. The large difference between these two data sources can inform the generalizability of the harm to female students' physics self-efficacy that Nissen and Shemwell identified. If there were no gender difference in the high-school data then Nissen and Shemwell's finding may have resulted from unique features of the specific course that they investigated. If, on the other hand, there were large gender differences in physics then this would indicate that the physics learning environment systemically harms female students' self-efficacy, particularly if the gender differences were unique to physics.

\section{The ESM Data Set}
The data I used for the investigation is a publicly available ESM data set collected in the 1990's as a part of the Sloan Survey of Youth and Social Development \citep{SSYSD} (SSYSD). The SSYSD data was collected in three waves in 1993, 1995, and 1997 from a total of 4,816 students at 33 schools in 12 locations that were selected to be representative of the United States as a whole. Data was collected using a broad range of instruments, but for my purpose I only focused on the ESM data. 
\par 	This ESM data was not explicitly collected to measure students' self-efficacy states. Therefore, I had to identify a self-efficacy state measure within the ESM data set. It was likely that the ESM data included a self-efficacy state measures since it included measures for students' feelings of skill, control, and success, which  formed the core of the self-efficacy state measure in Nissen and Shemwell's \citep{Nissen2016} earlier study.

\section{Methods} 
I used Principle Components Analysis (PCA) to identify the self-efficacy state construct in the SSYSD ESM data by checking if the skill, control, and success questions loaded on the same factor and identifying other questions that loaded on that factor. The PCA methods in this study matched those used by \citet{Nissen2016}. Once the component questions that formed the self-efficacy state measure were identified I averaged the component questions on the 5-point, 0-4, scale to match the scale used in Nissen and Shemwell's investigation.

\par 	I compared the means between male and female students' self-efficacy states in physics and in 9 other school activities. These school activities were categorized using the students' responses to the question about the main thing that they were doing as it was coded in the SSYSD data set. Cohen's \emph{d}, a measure of effect size, provided an easily comparable measure of the differences in the SSYSD data and the differences found in Nissen and Shemwell's earlier study of university students. T-tests with Bonferroni correction tested the reliability of the gender differences in self-efficacy states in each of the school activities. Bonferroni corrections minimized the possibility of false positive results that could have resulted from conducting several independent T-test by setting the alpha level for statistically reliable differences at \emph{p} \textless 0.005. I used histograms of male and female student's self-efficacy states in physics to compare the distributions of the two groups experiences and inform the meaningfulness of those differences. I included results from Nissen and Shemwell's \cite{Nissen2016} study in the results section to provide a context for any gender differences that are identified in the SSYSD data.

\section{Results}
Principle Components Analysis of 35,464 survey responses in the SSYSD data identified 6 factors with an eigenvalue greater than one. These six factors explained 58.7\% of the total variance in the data. The second factor had 6 components that loaded on it uniquely, including the three central components of the self-efficacy state measure (skill, control, and success), and explained 12.2\% of the variance on its own. Of the three core questions, success had an excellent loading (0.70), control had a very good loading (0.67) and skill had a very good loading (0.62). The strength of these loadings for these core questions supported interpreting this factor as a measure of self-efficacy states. Three other questions loaded on the same factor: ``Were you living up to your expectations?'' (0.76), ``Were you living up to others expectations?'' (0.67), and ``Did you feel good about yourself?'' (0.58). These three additional questions all aligned with the self-efficacy state being a measure of students feelings of ability and control in the activity at hand. The strength of the three core questions, the consistency of the additional three questions, and the variance explained by the factor support these 6 questions forming a reliable measure of self-efficacy states.

\begin{table}
\caption{Students' self-efficacy states.}
\begin{tabular}{p{1.9cm}cccp{.2cm}cccp{.2cm}c}
\hline \hline
&\multicolumn{3}{c}{\underline{   Male   }}&&\multicolumn{3}{c}{\underline{  Female  }}&&\\
Course 	&Mean	&N	&S.D.	&&Mean	&N	&S.D.	&&\emph{d}\\
\multicolumn{4}{l}{\emph{High School}}&&&&&&\\
Physics			&2.60	&48	&0.73	&&1.96	&21	&0.73	&&0.88*\\
Mathematics		&2.77	&364	&0.76	&&2.76	&518	&0.88	&&0.01\\
English			&2.82	&238	&0.74	&&2.77	&318	&0.82	&&0.06\\
Reading			&2.77	&114	&0.94	&&2.90	&161	&0.84	&&-0.15\\
Gen. Sci.			&2.73	&62	&0.82	&&2.85	&82	&0.92	&&-0.14\\
Biology			&2.70	&36	&0.75	&&2.69	&68	&0.65	&&0.00\\
Chemistry			&2.69	&31	&0.89	&&2.61	&50	&0.79	&&0.10\\
Comp. Sci.		&2.53	&58	&0.98	&&2.75	&63	&0.72	&&-0.26\\
For. Lang.			&2.67	&112	&0.70	&&2.84	&166	&0.81	&&-0.23\\
History			&2.64	&57	&0.82	&&2.71	&68	&0.83	&&-0.09\\
&&&&&&&&&\\
\multicolumn{4}{l}{\emph{College}}&&&&&&\\
Physics			&2.23	&148	&0.76	&&1.57	&82	&0.82	&&0.77*\\
STEM 			&2.41	&126	&0.71	&&2.25	&107	&0.73	&&0.22\\
Non-STEM 		&2.20	&99	&0.75	&&2.36	&62	&0.89	&&-0.20\\
\hline \hline
\multicolumn{4}{l}{* indicates \(\emph{p}<0.001\)}&&&&&&\\
\end{tabular}
\end{table}

\begin{figure}
\includegraphics[width=.5\textwidth]{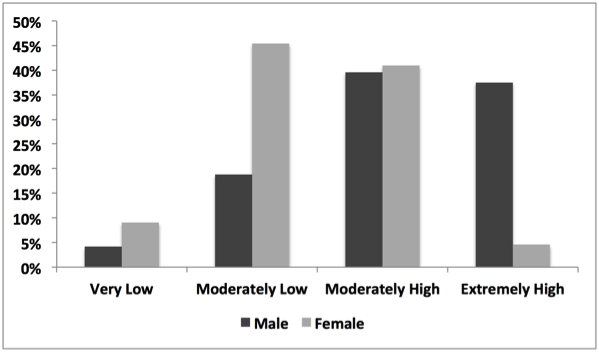}
\caption{Male and female students' self-efficacy states in high school physics.}
\end{figure}

\par 	Female students experienced numerically lower self-efficacy states in physics than male students did. This difference was more than three times larger than the next largest gender difference, as shown in Table I. The gender difference in students' self-efficacy states in physics was very large with an effect size of \emph{d} = 0.88 and was a statistically reliable difference with  \emph{p}\textless 0.001. This is slightly larger than the gender difference in a college physics course, which was \emph{d} = 0.77. These results indicated that there was something unique and reliable about female students experiencing very low self-efficacy states in the physics-learning environment.
\par 	The histograms of self-efficacy states in high school physics, Figure 1, indicated that male students largely experienced high levels of self-efficacy with 77\% of their self-efficacy states falling into the two high categories. Female students had a much lower proportion of self-efficacy states in the two high categories with the largest gender difference occurring in the very high category where males had 38\% of their experiences but females only had 4.5\% of their experiences. This was a very large difference. Concomitantly, female high school students were twice as likely to have very low self-efficacy states in physics than male students and twice as likely to have moderately low self-efficacy states. These distributions show that female students were much less likely to have the very high self-efficacy states that could support their development of self-efficacy traits and much more likely to have the very low self-efficacy states that could harm their self-efficacy traits.

\par	The size of the gender difference in self-efficacy states in physics was similar in both high school and university courses. However, male and female students' self-efficacy states in high school physics were higher than their counterparts in university physics courses. The difference in self-efficacy states between high school and university were not unique to physics as the means in self-efficacy states in high school classes tended to be about 0.4 to 0.6 higher on the 0 to 4 scale than university courses with similar content.

\section{Discussion}
The three core components of the self-efficacy state (skill, control, and success) formed a single, consistent and reliable construct in high school students' daily experience. Combined with the similar finding of \citet{Nissen2016} this result indicates that self-efficacy is a consistent component of students' daily experiences. Given the established relationships between self-efficacy and student outcomes the ability to measure self-efficacy in the midst of students' experiences that the ESM provides can be used to further inform the relationships between self-efficacy, learning, and performance.
\par 	Female students experienced much lower self-efficacy states in high school physics courses than their male peers and this difference was very large. They were much more likely to experience very low self-efficacy states and much less likely to experience very high self-efficacy states. This distribution of experiences was consistent with an environment that harms self-efficacy and was strikingly similar to the distribution of female college students' self-efficacy states in physics \citep{Nissen2016}. The large gender difference in self-efficacy states in physics, and only in physics, from a representative sample of high school students in combination with a similarly large gender difference in a university physics course indicates that most, if not all, physics instruction tends to harm female students' physics self-efficacy. This general harm is further evidenced by female students' self-efficacy traits decreasing across many forms of university physics instruction, including both research-based instruction and lecture-based instruction.

\section{Conclusion}
Populations of students that regularly experience little to no success or skill in physics are not likely to be a source of many aspiring physicists, particularly if those students also report little to no control over their experiences despite achieving similar course grades to other students \cite{Nissen2016}. Currently, physics instruction systemically harms female students self-efficacy. Changing that status quo and supporting female students in developing physics self-efficacy can lead more, and more diverse, students to pursue and succeed in physics careers. A starting point for this change is students' entry point into physics education at either the high school or collegiate level. Supporting students' self-efficacy at this early point will support those students in developing the cognitive and affective traits that will increase their likelihood of pursuing a career in physics and their ability to succeed in that pursuit.

\section{Directions for Research}
This investigation is part of ongoing research on the sources of the harm to female students physics self-efficacy that is focused on the relationships between self-efficacy and stereotype threat. Stereotype threat is ``a situation where one faces judgment based on societal stereotypes about one's group'' \citet[p.~5]{Spencer1999}. Similar to self-efficacy, stereotype threat is a focus for education researchers because it negatively impacts students' test performance \cite{Steele1995, Sekaquaptewa2003} and ability to learn \cite{Rydell2010}. Stereotype threat and low self-efficacy states are also both related to stress \citep{Nissen2016,Mangels2012,Osborne2007}, and, similar to results in this study, stereotype threat occurs in physics but not in chemistry \cite{Deemer2014,Marchand2013}. Investigating the relationships between self-efficacy and stereotype threat can lead to a greater understanding of both of these phenomena \cite{Usher2008} that will support educators and researchers in understanding and addressing complex social problems that extend beyond equity in physics. 


\bibliography{gender}

\end{document}